\documentclass[pra,twocolumn,aps,showpacs,superscriptaddress,longbibliography]{revtex4-2} 

\usepackage{amsmath}
\usepackage{amssymb}
\usepackage{amsmath}
\usepackage{txfonts}
\usepackage{graphicx} 
\usepackage{epsfig}
\usepackage{float} 
\usepackage[T1]{fontenc}
\usepackage{times}
\usepackage{color} 
\usepackage{cases}
\usepackage{xspace}
\usepackage{amssymb,amsmath}
\usepackage{amsbsy}
\usepackage{braket}
\usepackage{tabularx}
\usepackage{mathtools}  

\usepackage[utf8]{inputenc}

\DeclareMathOperator{\sech}{sech}

\usepackage{ifpdf}
\ifpdf
\usepackage{epstopdf}
\fi

\usepackage[unicode]{hyperref}
\hypersetup{
	pdftitle={spinnorBEC},
	pdfauthor={Yu},
	pdfsubject={Vortexlines},
	colorlinks=true,
	linkcolor=blue,
	citecolor=blue,
	filecolor=black,
	urlcolor=blue
}
\def\urlprefix{}
\def\url#1{}

\usepackage{bm}
\usepackage{color}

\newcommand{\be}{\begin{equation}}
	
	\newcommand{\ee}{\end{equation}}

\newcommand{\bea}{\begin{eqnarray}}
	
	\newcommand{\eea}{\end{eqnarray}}

\DeclareUnicodeCharacter{2212}{\textendash}

\begin{document}

	\title{Ferrodark soliton collisions:  Breather formation, pair reproduction, and spin-mass separation } 
	
	\author{Yixiu Bai}
	
	\affiliation{Graduate School of China Academy of Engineering Physics, Beijing 100193, China}

	\author{Jiangnan Biguo}
	\affiliation{Graduate School of China Academy of Engineering Physics, Beijing 100193, China}
	
	\author{Xiaoquan Yu}
	\email{xqyu@gscaep.ac.cn}
	\affiliation{Graduate School of China Academy of Engineering Physics, Beijing 100193, China}
	\affiliation{Department of Physics and Centre for Quantum Science, University of Otago, Dunedin 9016, New Zealand}

	\begin{abstract}
		We study collisions between a ferrodark soliton (FDS) and an antiFDS ($\mathbb{Z}_2$ kinks in the spin order) in the easy-plane phase of spin-1 Bose-Einstein condensates (BECs). 
		For a type-I pair (type-I FDS-antiFDS pair) at low incoming velocities, the pair annihilates followed by the formation of an extremely long-lived dissipative breather on a stable background,  a spatially localized wave packet with out-of-phase oscillating magnetization and mass superfluid densities. Periodic emissions of spin and density waves cause breather energy dissipation  and we find that the breather energy decays logarithmically in time.  When the incoming velocity is larger  than a critical velocity at which a stationary FDS-antiFDS pair forms,  a pair with finite separating velocity is reproduced.  When approaching the critical velocity from below, we find that the lifetime of the stationary type-I pair shows a power-law divergence, resembling a critical behavior. In contrast, a type-II pair (type-II FDS-antiFDS pair) never annihilates and only exhibits reflection.  For collisions of a mixed type FDS-antiFDS pair,  as $\mathbb{Z}_2$ kinks in the spin order, reflection occurs in the topological structure of the magnetization while the mass superfluid density profiles pass through each other,  manifesting spin-mass separation.  
	\end{abstract}
	
	\maketitle
\section{Introduction} 
Kinks and domain walls are topological solitons, i.e., solitons with topological characteristics, which exist in many condensed matter systems~\cite{chaikin1995principles, nelson2002defects} and are expected to  play important roles in the evolution of early universe~\cite{Kinksanddomanwalls}.  Collisions among topological solitons reveal the interaction between them and are  crucial in determining the universality of relevant nonequilibrium processes~\cite{KZspin12007,Gasenzer18,Fujimoto19}. 
	
	Solitons in classical wave systems described by the Korteweg-de Vries (KdV) equation \cite{KdV,Drazin_Johnson_1989}  exhibit shape-preserving collision dynamics,  a property  regarded as a defining property of solitons and a feature of integrable systems.  Similar properties also hold for kinks in integrable models.  The representative examples are sine-Gordon kinks and in collisions two kinks pass through each other without energy dissipation~\cite{Kinksanddomanwalls,Sickotra21,Drazin_Johnson_1989}. 
	Collisions of kinks in nonintegrable systems show more complex characters~\cite{RMP1989,Kinksanddomanwalls}.  For $\mathbb{Z}_2$ kinks in the $\phi^4$  model, the outgoing state of head-on collisions of a kink and antikink pair depends on their initial velocities~\cite{Kinksanddomanwalls,Campbell83}.  
	In superfluids, propagating kinks are rare, so kink collisions remain poorly understood. Recently discovered ferrodark solitons (FDSs) in spin-1 BECs are propagating $\mathbb{Z}_2$ kinks in the spin  order~\cite{Xiaoquan21,Xiaoquan22,Xiaoquan22Corestructure, FDSsnake2024,FDSspincorrection} and provide a unique example to study rich dynamics of kink collisions in superfluids.  
	
	In this work, we numerically investigate collisions of FDS-antiFDS pairs in a uniform ferromagnetic spin-1 BEC.  At low initial velocities,  a type-I pair annihilates accompanied by the formation of a breather characterized by spatially localized periodic motion of magnetization and mass superfluid densities. Such a magnetic breather is extremely  long-lived and persists in our longest  time numerical simulations. Moreover, we find that the magnetic breather formed through the scattering process  dissipates energy  via emitting density and spin waves periodically and the breather energy shows logarithmic decay.  When the incoming velocity is high enough, a type-I pair with finite separating velocity is reproduced. Then there exists a critical velocity, at which a stationary type-I pair is reproduced.  We find that when approaching the critical velocity from below, the stationary type-I pair has a power-law divergent lifetime.   
	In contrast, a type-II pair only exhibits reflection and no annihilation occurs. For collisions between a type-I FDS and a type-II antiFDS, they reflect in the spin order as $\mathbb{Z}_2$ kinks,  however, they pass through each other in the mass superfluid density, showing a phenomenon of spin-mass separation. Moreover,  for the pair positioned symmetrically with respect to the origin,  the collision center is shifted to the side of the type-I FDS,  caused by the in-elasticity of the collision and the distinct dispersion relations of the type-I and type-II FDSs.  
	

\section{Systems and FDSs}

In an optical trap, the spin degrees of freedom of atoms are liberated~\cite{Stenger98_exp,Stemper-Kurn_optical_trap_exp}, allowing exchange dynamics  between different spin states. A spin-1 BEC is a condensate consisting of atoms in hyperfine states $|F=1,m=-1,0,+1\rangle$, where $F$ denotes the total spin quantum number.  In the mean-field level, the Hamiltonian density of a spin-1 BEC reads~\cite{Ho98,Ohmi98,Kawaguchi12,StamperKurn13}:
	\begin{equation}
		H=\frac{\hbar^2}{2M}|\nabla\psi|^2+\frac{g_n}{2}|\psi^\dagger\psi|^2+\frac{g_s}{2}|\psi^\dagger{\rm \bf S}\psi|^2+q\psi^\dagger S^2_z\psi,
	\end{equation}
	where $\psi=(\psi_{+1},\psi_0,\psi_{-1})^T$ is  the three-component  wave function, $M$ is the atomic mass,  $g_n$ is the density-density interaction strength,  $g_s$ is the spin-dependent interaction strength,  ${\rm \bf S}=(S_x,S_y,S_z)$ with $S_{i=x,y,z}$ being spin -1 matrices~\cite{footnoteSO(3)}, and $q$ is the quadratic Zeeman energy. The magnetization ${\bf F}\equiv\psi^\dagger{\bf S}\psi$ serves as the order parameter. For the ferromagnetic coupling ($g_s<0$) and the choice $F_z=0$, the ground state isthe easy-plane phase when $0<\tilde{q}\equiv q/(-2g_s n_b)<1$~\cite{StamperKurn13,Kawaguchi12}, characterized by $F_\perp \equiv F_x + iF_y \neq0$, where $n_b$ is the ground state total number density.The evolution of $\psi$ is governed  by the spin-1 Gross-Pitaevskii equations (GPEs) :
	\begin{eqnarray}
		i\hbar \frac{\partial \psi_{\pm1}}{\partial t}&=&\left[H_0+g_s(n_0+n_{\pm1}-n_{\mp1})+q\right]\psi_{\pm1}+g_s\psi_0^2\psi_{\mp1}^*, \\
		i\hbar\frac{\partial\psi_0}{\partial t}&=&\left[H_0+g_s(n_{1}+n_{-1})\right]\psi_0+2g_s\psi_0^*\psi_{1}\psi_{-1},
	\end{eqnarray}
	where $H_0=-\hbar^2\nabla^2/2M+g_n n$,  $n=\sum_{m} n_m$ is the total number density and $n_m=\psi_m^\dagger\psi_m$ is the component density.


	In the easy-plane phase,  there exists $\mathbb{Z}_2$ magnetic kinks in the spin order and they are the FDSs~\cite{Xiaoquan21,Xiaoquan22,Xiaoquan22Corestructure, FDSsnake2024,FDSspincorrection}. The FDSs have two types and type-I (II) FDSs have positive (negative) inertial mass. At $g_s/g_n=-1/2$ (close to the value for $^7$Li~\cite{Huh2020a}), the exact wave functions of propagating FDSs read: 
	\bea 
	\psi^{\rm I}_{\pm1}&=&\sqrt{n_b^{\pm1}}\left(-\alpha\tanh \zeta^{\rm I}+i\delta\right),\\ \psi^{\rm I}_{0}&=&\sqrt{n_b^{0}}\left(\alpha-i\delta\tanh\zeta^{\rm I}\right), \\
	\psi^{\rm II}_{\pm1}&=&\sqrt{n_b^{\pm1}}\left(\beta+i\kappa\tanh\zeta^{\rm II}\right),\\  \psi^{\rm II}_{0}&=&\sqrt{n_b^{0}}\left(-\beta\tanh \zeta^{\rm II}-i\kappa\right),
	\eea
	 where $\zeta^{\rm I,II}=(x-V t)/\ell^{\rm I,II}$, $V$ is the propagating velocity, $\ell^{\rm I,II}=\sqrt{2\hbar^2/[M(g_nn_b-MV^2\mp Q)]}$ characterizes the soliton width with the minus (plus) sign in front of Q specifying type-I (II) FDSs, $Q=\sqrt{M^2V^4+q^2-2g_nn_bMV^2}$, $n_b^{0,\pm1}$ are ground state component densities, and
	 \bea
	 \alpha&=&\sqrt{\frac{q+MV^2+Q}{2q}}, \\
	 \delta&=&\sqrt{\frac{q-MV^2-Q}{2q}}, \\
	\beta&=&\sqrt{\frac{q-MV^2+Q}{2q}}, \\
	\kappa&=&\sqrt{\frac{q+MV^2-Q}{2q}}.
	\eea
Note that here the coefficients  $\alpha$, $\delta$, $\beta$ and $\kappa$ are considerably simplified comparing with their original expressions reported in Ref.~\cite{Xiaoquan22}.  At the speed limit  $V=c_{\rm FDS}=\sqrt{g_nn_b/M}\sqrt{1-\sqrt{1-(q/g_n n_b)^2}}$, $Q=0$ and two types FDSs become identical.  The FDS energy 
	\bea \delta K^{\rm I,II}(q,V^2)=\frac{\sqrt{2}\hbar (g_nn_b-MV^2\mp Q)^{3/2}}{3 g_n\sqrt{M}}
	\eea
	increases (decreases) with increasing the velocity $V$ for type-I (II) FDSs and $\delta K^{\rm II}(q,V^2)>\delta K^{\rm I}(q,V^2)$. 
	The transverse magnetization  and the superfluid density are~\cite{Xiaoquan22}:
	\bea
	 F_{\perp}^{\rm I, II}(x,t)&=&-\sqrt{n_b^2-\frac{q^2}{g_n^2}}\tanh\left(\frac{x-Vt}{\ell^{\rm I,II}}\right), \\
	  n^{\rm I, II}(x,t)&=&n_b-\frac{g_nn_b-MV^2\mp Q}{2g_n}\sech^2\left(\frac{x-Vt}{\ell^{\rm I,II}}\right).
	 \eea

	%



		\begin{figure}[htbp]
	\centering
	\hspace*{-0.6cm} 
	\includegraphics[width=\linewidth]{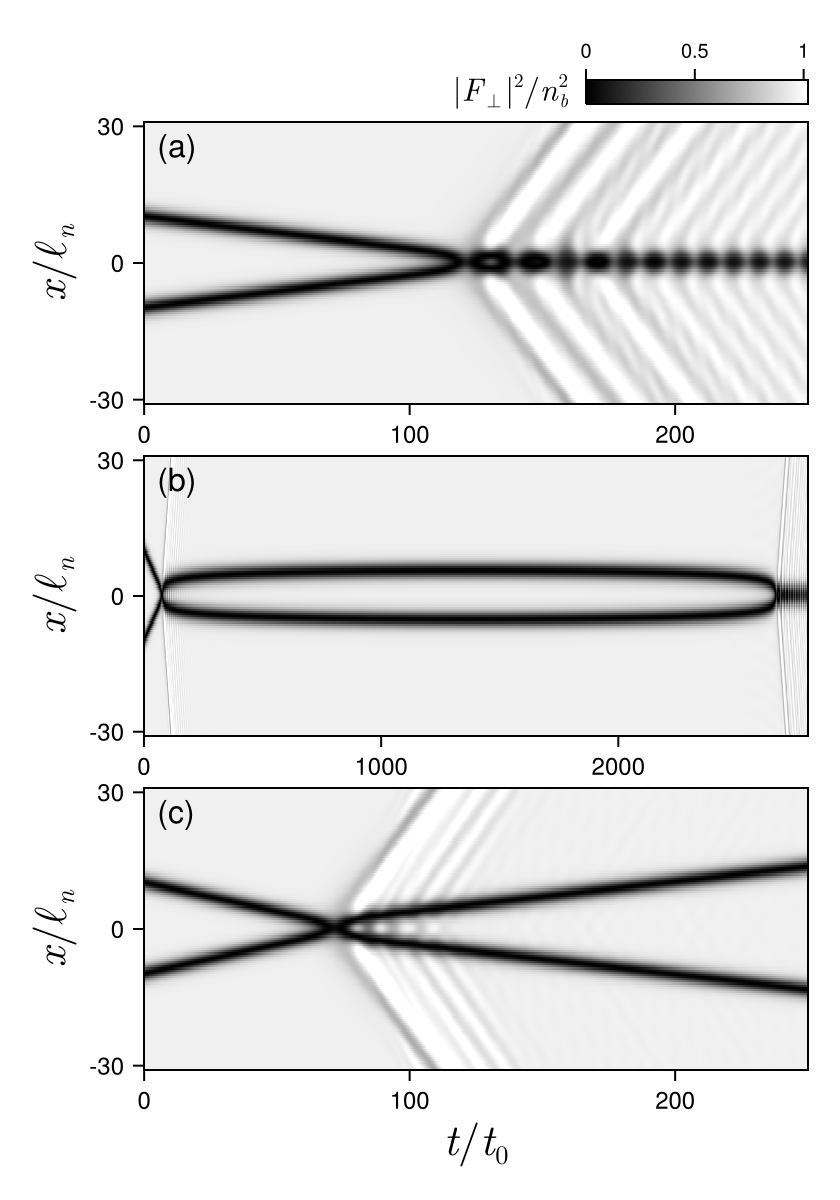}
	\caption{Collisions between a type-I FDS and a counterpropagating type-I antiFDS (a type-I pair) with incoming velocities: (a) $V_\mathrm{in}/c_{\rm{FDS}}=0.5$, (b) $V_\mathrm{in}/c_{\rm{FDS}}=0.846 25$ , and (c) $V_\mathrm{in}/c_{\rm{FDS}}=0.9$.  (a)  A type-I pair annihilates at low velocities followed by the formation of a breather.  (b) A stationary type-I pair is reproduced near a critical incoming velocity. The pair eventually annihilates driven by the weak overlap between the two FDSs. (c) Reproduction of a propagating type-I pair at high velocities. Here $t_0=\hbar/g_nn_b$, $\ell_n=\sqrt{2\hbar^2/Mg_nn_b}$, $g_s/g_n=-0.5$ and $\tilde{q}=0.2$. } 
	\label{collision_processes}
\end{figure}

\begin{figure*}[htp]
	\centering
	\hspace*{-0.6cm} 
	\includegraphics[width=\linewidth]{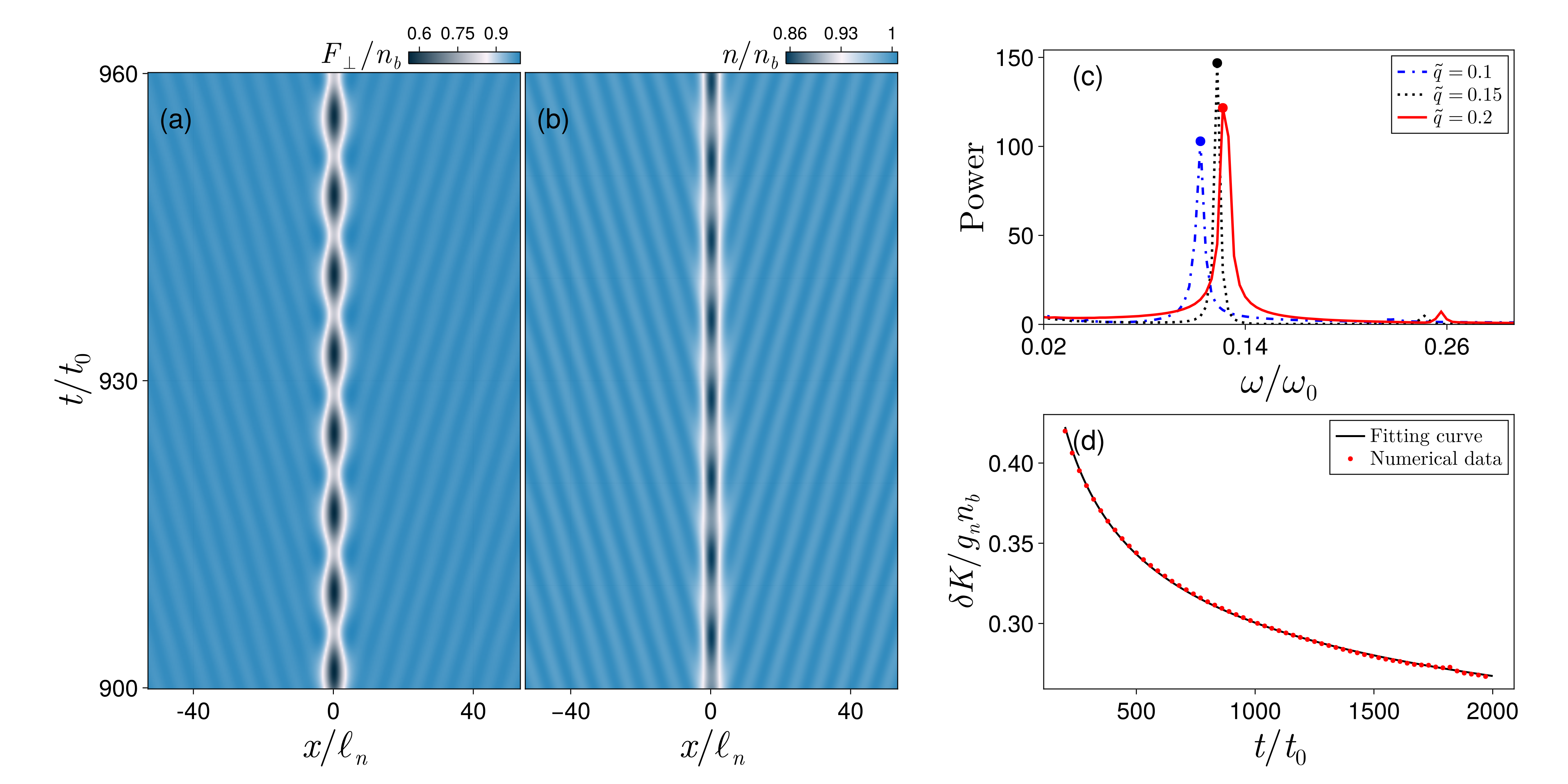}
	\caption{The time evolution of the magnetic breather formed after a type-I pair annihilation at  $V_\mathrm{in}/c_\mathrm{FDS}=0.5$. The formed breather is characterized by the out-of-phase oscillations of the magnetization density (a) and the total number density (b) accompanied by the periodically emitting spin and number density waves. Here the time window is $t/t_0: 900-960$ in which the breather is well-formed after the annihilation. (c) The power spectrum of the time evolution of $F_{\perp}$ at $x=0$ within the time window $t/t_0: 700-1000$, where the peaks indicate breather oscillation frequencies for different values of $\tilde{q}$.  (d) The breather energy decays logarithmically.  Here the fitting function is $\delta K/g_n n_b=A/[\log (t/t_0)+B]$, where the best fitting parameters $A \simeq 1.675$ and $B \simeq -3.636$. The breather energy $\delta K$ is less than the energy of the stationary type-I pair $2\delta K^{\rm I}(V^2=0)/(g_n n_b)\simeq 0.675$ as it should be. Here $\omega_0=1/t_0$, $g_s/g_n=-0.5$ and $\tilde{q}=0.2$.}
	\label{fig:breather}
\end{figure*}

\begin{figure*}
	\centering
	\includegraphics[width=1\textwidth]{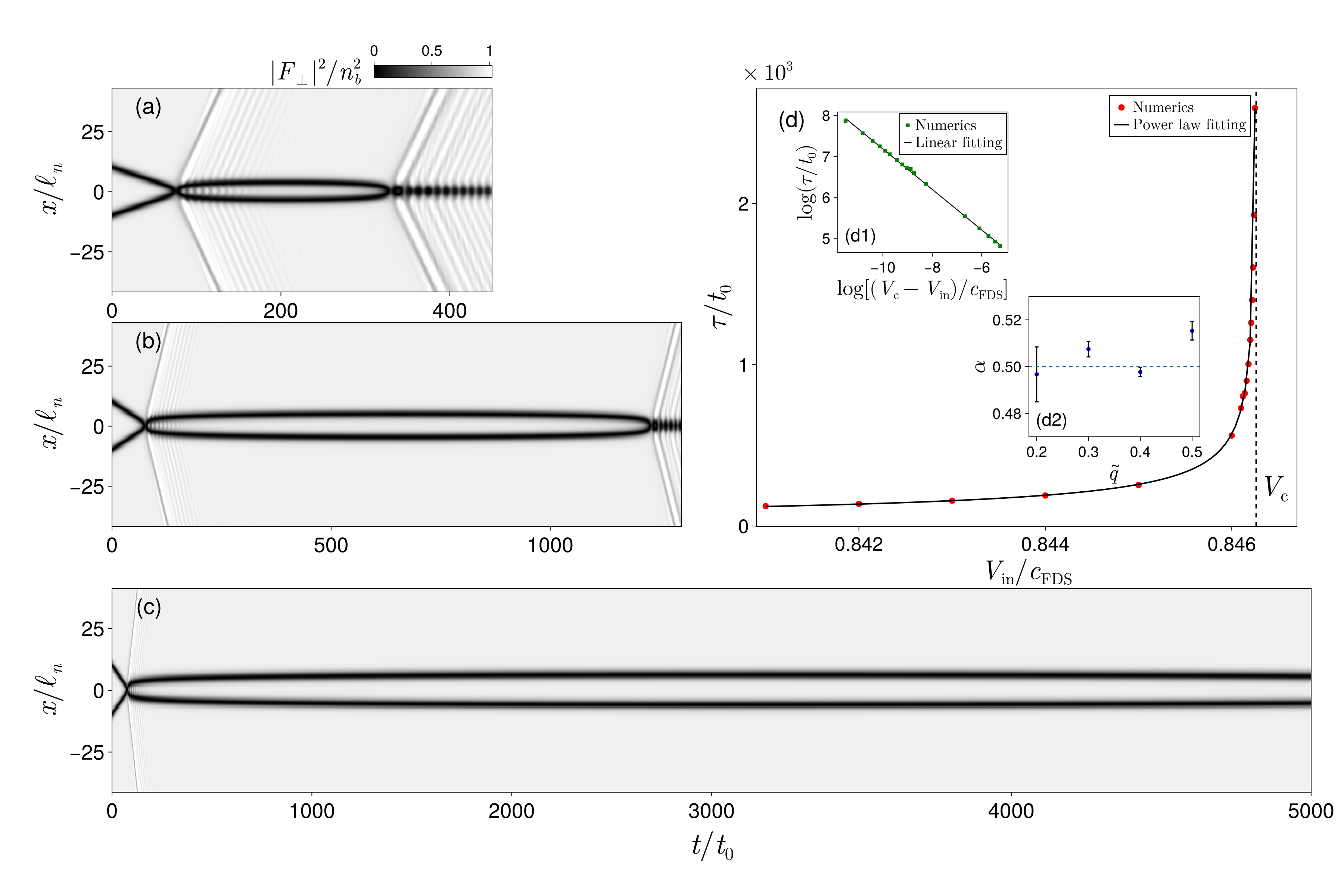}
	\caption{Collisions of a type-I pair for incoming velocities gradually approaching the critical incoming velocity:  $V_\mathrm{in}/c_\mathrm{FDS}=0.845$(a),  $V_\mathrm{in}/c_\mathrm{FDS}=0.8462$(b) and $V_\mathrm{in}/c_\mathrm{FDS}=V_c/c_\mathrm{FDS}\simeq0.84626$ (c). Here $\tilde{q}=0.2$ and $g_s/g_n=-0.5$, which are the same as the parameters adopted in Fig.~\ref{collision_processes}.  The lifetime of the reproduced stationary type-I pair as function of the incoming velocity is shown in (d), where the fit function is $\tau/t_0=A [(V_\mathrm{c}-V_\mathrm{in})/c_{\rm FDS}]^{-\alpha}$ with the best fitting parameters $\alpha \simeq 0.4967$,$V_ {\mathrm{c}}/c_{\rm FDS}\simeq0.84626$, and $A\simeq9.44$.   The power law behavior is confirmed by showing a linear behavior of the data on the logarithmic scale [inset (d1)].  The exponent $\alpha$ is weakly dependent on $\tilde{q}$ and takes the value around $1/2$ [inset (d2)]. }
	\label{fig:pairlifetime}
\end{figure*}	
\begin{figure}
	\centering
	\hspace*{-0.6cm} 
	\includegraphics[width=\linewidth]{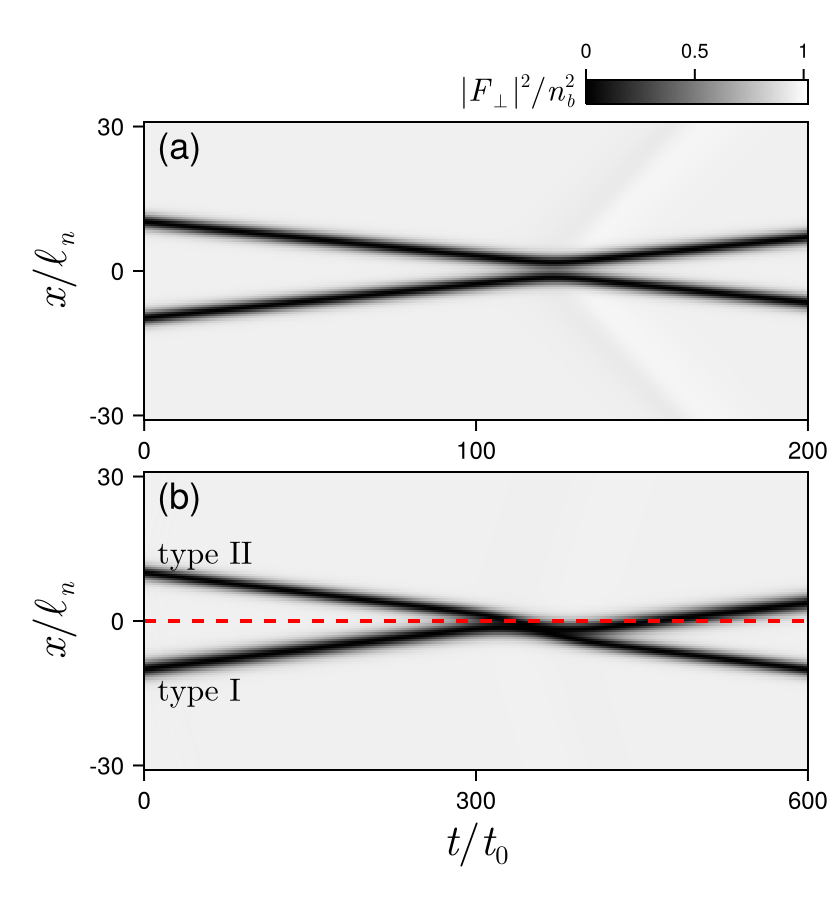}
	\caption{Collisions between a FDS and a counterpropagating antiFDS: a type-II pair (a) and a mixed type pair (b) with the incoming velocity $V_\mathrm{in}/c_{\rm FDS}=0.5$ and $V_\mathrm{in}/c_{\rm FDS}=0.2$, respectively.  (a) Reflection of a type-II pair without annihilation. (b) Collision between a FDS-antiFDS pair of mixed types with a shift of the collision center. Here $g_s/g_n=-0.5$ and $\tilde{q}=0.2$. }
	\label{fig:II_mix}
\end{figure}

\begin{figure*}
	\centering
	\includegraphics[width=0.98\textwidth]{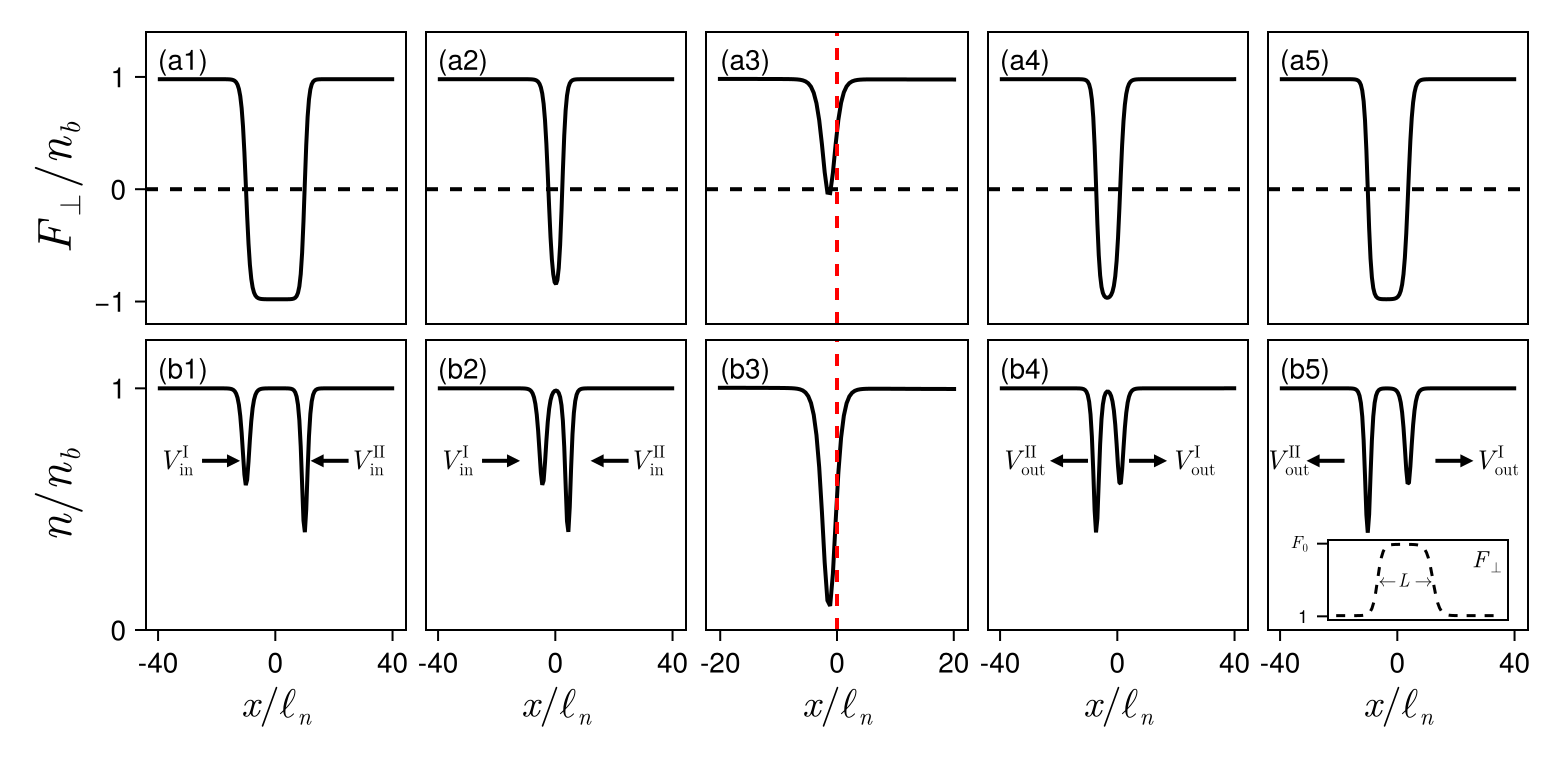}
	\caption{{The profiles of the transverse magnetization density $F_{\perp}$ [(a1)-(a5)] and the mass superfluid density  $n$ [(b1)-(b5)] during the collision of the pair composed by a type-I FDS and a type-II anti-FDS described in Fig.~\ref{fig:II_mix} (b) at $t/t_0=1;270;350;500;600$.  Reflection occurs  in the spin order [(a1)-(a5)], while the mass superfluid  density clearly shows passing through [(b1)-(b5)]. The inset in (b5) shows the profile of  $F_{\perp}$ if passing through in the spin order happened. }}
	\label{fig:differenttype}
\end{figure*}

\section{Collisions of a type-I pair: formation of magnetic breathers and pair reproduction}
 Let us consider  collisions between a FDS and a counterpropagating antiFDS of the same type in a uniform system.  Note that  two $\mathbb{Z}_2$ kinks can not be adjacent to each other and  hence only kink-antikink scattering is possible~\cite{Kinksanddomanwalls}. 
	We prepare a largely-separated  type-I pair with opposite velocities. The pair firstly annihilates, accompanied by the destruction of the kink structure in $F_\perp$  and the out-going state depends on the incoming velocity.  At low incoming  velocities,   the energy loss (via emitting density and spin waves) is large after the first collision and hence the reproduction of a new type-I pair is forbidden.  Strikingly,  the type-I pair does not annihilate completely into linear waves but forms a long-lived spatially localized  nonlinear wave packet for moderate  values of $\tilde{q}$---a monochromatic breather [Fig.~\ref{collision_processes}(a)], characterized by oscillations of  the  magnetization $F_{\perp}$ and the mass superfluid density $n$ [Fig.~\ref{fig:breather}(a)(b)]  with a frequency  weakly depending on the quadratic Zeeman energy and the incoming velocity [Fig.~\ref{fig:breather}(c)].  The formed breather is weakly dissipative via periodically radiating density and spin waves and its energy decays logarithmically  [Fig.~\ref{fig:breather}(d)], ensuring its extremely long lifetime (Fig.~\ref{breatherlongtime} in Appendix~\ref{breatherlonglived}).	A similar energy decay behavior was also observed for the breather-like excitation formed after the collision of a $\mathbb{Z}_2$ kink-antikink pair in the $\phi^4$ model~\cite{logdecay1994}.  Here we identify that  the emitted waves by the formed breather belong to the gapless Bogoliubov mode with higher energy and the corresponding density wave and spin wave propagate at the same velocity [see Fig.~\ref{densitywaves} in Appendix~\ref{Density waves}].
    We should emphasize the observed weakly dissipative breather is on a stable background and  is clearly distinct from well-known  
	breathers such as Akhmediev and  Kuznetsov-Ma breathers which are solutions of an integrable system and  on an unstable background~\cite{akhmediev1986modulation, kuznetsov1977solitons, ma1979perturbed}. 
	Moreover, to the best of our knowledge, the only known models which support the phenomenon of fusion of a kink-antikink pair into a breather are  the $\phi^4$ model~\cite{Kinksanddomanwalls,Campbell83} and the perturbed  sine-Gordon model~\cite{MALOMED1985,RMP1989}.  Here we show that spin-1 BECs support similar phenomenon, providing a new platform to investigate the relevant rich dynamics. When the incoming velocity reaches a threshold,  a stationary type-I pair is reproduced  after the annihilation [Fig.\ref{collision_processes}(b)].  The numerical results  suggest that there exists a critical incoming velocity at which a true stationary type-I pair is formed signed by the divergence of its lifetime [Fig.~\ref{fig:pairlifetime}(a)-(c)]. It turns out that the lifetime of the stationary type-I pair exhibits a power-law divergence, resembling a critical behavior [Fig.~\ref{fig:pairlifetime}(d)].

At sufficiently high initial velocities,  a new type-I pair can be recreated with a finite separating velocity [Fig.\ref{collision_processes}(c)], which can be effectively viewed as reflection.  
	The conspicuous feature here is that a type-I pair always annihilates firstly when colliding. If the remaining energy of the localized nonlinear wave packet is less than the energy of a stationary type-I pair,  a reproduction can not occur, which is the case for low incoming velocities (low initial energies).  For a type-I pair with high incoming velocities (high initial energies), the remaining energy is sufficiently large and recreating  a separating  type-I pair is possible~[Fig.~\ref{collision_processes}(c)][see also Fig.~\ref{fig:kinkstructure}(a2)(a3)(a4) in Appendix ~\ref{kink structure}],  where the separating  velocity is lower than the incoming velocity since the type-I FDS energy increases as increasing the propagating velocity~\cite{Xiaoquan22}.   
	Here we emphasize that the collision dynamics of type-I pairs investigated in the exactly solvable parameter region holds qualitatively for other values of $g_s/g_n$ (Fig.~\ref{gs001}(a)-(c) in Appendix~\ref{away solvable regime}).

It is useful to summarize the similarities and the differences between collisions of $\mathbb{Z}_2$ kink-antikink pair in the $\phi^4$ model and collisions of type-I FDS-antiFDS (kinks in the spin order) pairs.  The ferromagnetic spin-1 GP model and the $\phi^4$ model are not integrable and collisions for both kink-antikink pairs are inelastic. At very low incoming velocities, the $\mathbb{Z}_2$ kinks annihilate followed by the formation of  a breather-like excitation (or a bound state of the kink-antikink pair).  At very high incoming velocities, reflection happens.  For the intermediate incoming velocities there exists bands of incoming velocity at which annihilation and reflection takes place alternatively~\cite{Kinksanddomanwalls,Campbell83}. 
Similarly, a breather-like excitation  also forms for a type-I pair at low velocities. While here the breather has richer structure characterized by out-of-phase oscillations of  magnetization and number densities.  For both systems, the formed breathers are dissipative and their energies decay logarithmically in time. 
 At very high incoming velocities, reflection or more precisely the reproduction of a separating pair takes place.  Different from the $\mathbb{Z}_2$ kink collisions in the $\phi^4$ model, for a type-I pair the change from annihilation to reflection happens at one critical value of the incoming velocity.    An another distinct feature which was not observed in the $\phi^4$ model is the formation of a stationary pair at the critical incoming velocity and the power-law divergence of its lifetime near the critical incoming velocity.

	
\section{Collisions of a type-II pair and a mixed type pair: reflection and spin-mass separation}
In contrast to type-I pairs, during  the collision of a type-II pair  [Fig.\ref{fig:II_mix}(a)]  the kink structure is preserved and the pair does not involve annihilation (kink structure destruction) [{Fig. ~\ref{fig:kinkstructure} (a1')-(a5') in Appendix~\ref{kink structure}]. The energy loss is still present (however is much weaker in comparison with collisions of a type-I pair), yielding that the reflected velocity is higher than the incoming velocity since the type-II FDS energy decreases as increasing the propagating velocity~\cite{Xiaoquan22}. Again, the collisions of type-II pairs investigated in the exactly solvable parameter region holds qualitatively for other values of $g_s/g_n$ (Fig.~\ref{gs001} (d)-(f) in Appendix~\ref{away solvable regime}).
	
	We now consider a FDS and an antiFDS of a different type with opposite velocities and initial positions symmetric with respect to the origin.  
	As $\mathbb{Z}_2$ kinks in the spin order,  the type-I FDS and the type-II antiFDS reflect.  This is a generic feature of  $\mathbb{Z}_2$ kinks~\cite{Campbell83} since if passing through happens,  the region between the kink and the antikink is at some value $F_0$ different from the ground state value,  the energy of this configuration is proportional to $L$ [inset in Fig.~\ref{fig:differenttype} (b5)] 
	and  hence  $L $ can not approach infinity. However in viewing of the mass superfluid density the type-I FDS and the type-II antiFDS pass through each other [Fig.~\ref{fig:differenttype}(b1)-(b5)].  This phenomenon manifests spin-mass separation, a novel feature in FDS collisions due to the presence of two orders in the spinor superfluid. 
	Yet, it is not clear the reason why the mass superfluid densities prefer passing through rather than reflecting, deserving future investigations. 
	 A related phenomenon called spin-charge separation is normally  expected to occur in strongly correlated systems~\cite{giamarchi2003quantum}.  
Another notable  feature is that  the collision position is not at the origin but is shifted towards the incoming side of  the type-I FDS [Fig.~\ref{fig:II_mix}(b), Fig.~\ref{fig:differenttype} (a3)(b3)].  
	For the inelastic collision process, the energy loss is unavoidable and therefore  the out-going velocity of the type-I FDS decreases while the out-going velocity of the type-II antiFDS increases~\cite{Xiaoquan22},  yielding the  shift of the collision position.  No annihilation appears in this collision process [Fig.~\ref{fig:differenttype}(a3)]. 
	
	
\section{Conclusion} 
We numerically investigate inelastic collisions of  FDS-antiFDS pairs  in ferromagnetic spin-1 BECs.  We find that a type-I pair annihilates (destruction of the topological structure of  kinks in the spin order) at low incoming velocities followed by the formation of an extremely long lived dissipative breather featured by the oscillations of spatially localized magnetization and number densities. The breather periodically emits spin and density waves, causing a logarithmic energy decay. When approaching  the critical incoming velocity, the lifetime of a  formed stationary pair after the annihilation shows a power-law divergence.  At velocities higher than the critical incoming velocity,  the outgoing state is a type-I pair with a finite separating velocity. For type-II pairs, annihilation does not happen and only reflection occurs.  A mixed type-I type-II pair exhibits spin-mass separation, i.e., reflection in the spin order (as a $\mathbb{Z}_2$ kink) and passing through in the mass superfluid density.  Note that  transitions between type-I FDSs and type-II FDSs may occur during the scattering processes for very small values of the quadratic Zeeman energy  and the relevant phenomena will be discussed elsewhere. This work explores the new regime of  kink collision dynamics and will also motivate experimental investigations~\cite{Dalibard2015,Gauthier16,Semeghini2018,Higbie2005,Huh2020a,MSexp1, MSexp2, prufer2022condensation}.  In particular, our results might be useful for interpreting the experimentally observed  scaling laws in quench dynamics of spin-1 BECs~\cite{Gasenzer18}, where ferrodark solitons are possibly the relevant topological defects.

\section{Acknowledgment} 
	We thank C. Ma, R. Han, P. Ao, H. Pu, J. Wang and H. Hu for useful discussions. X.Y. acknowledges support from the National Natural Science Foundation of China (Grant No. 12175215, Grant No. 12475041), the National Key Research and Development Program of China (Grant No. 2022YFA 1405302) and  NSAF (Grant No. U2330401).


\appendix
\onecolumngrid 

\newpage

\section{Long-lived breathers}	\label{breatherlonglived}
As claimed in the main manuscript, the breather emerged from the annihilation of a type-I pair has long lifetime.  Here we present  long time simulations and indeed the breather is still present  long after its formation (Fig.~\ref{breatherlongtime}).

\begin{figure} [h]
	\centering
	\includegraphics[width=0.8\linewidth]{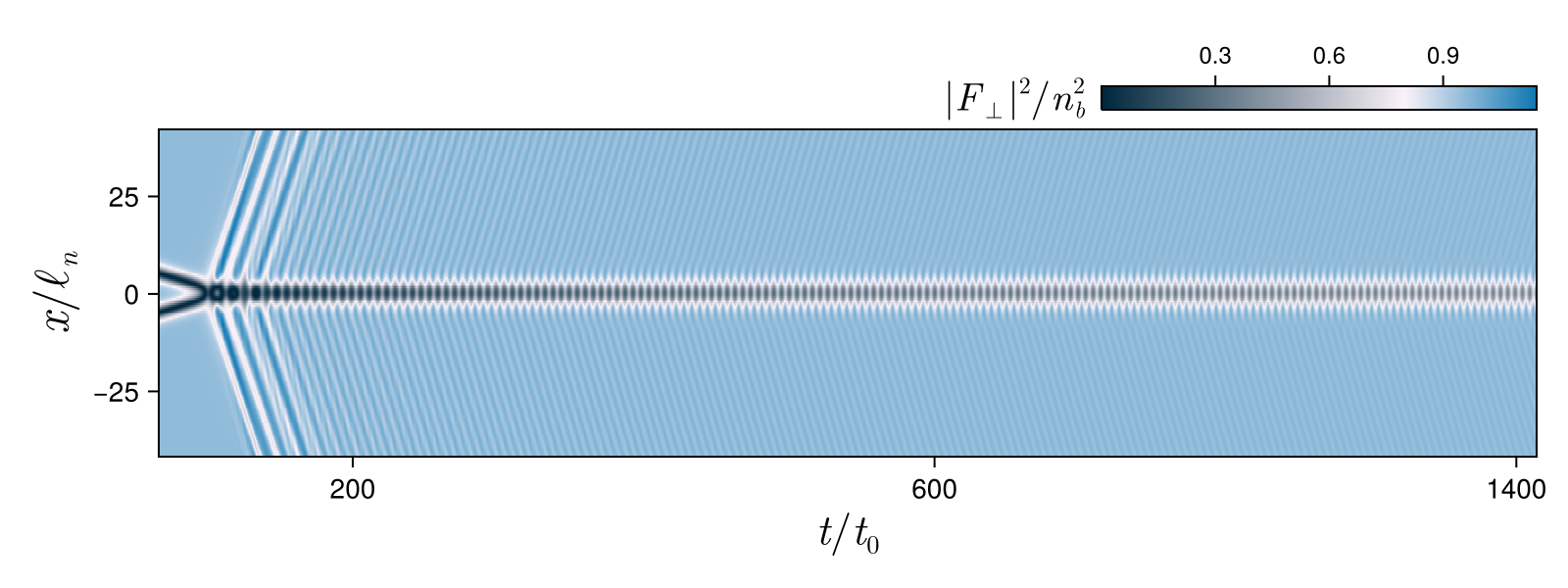}
	\caption{Long time evolution of the breather after its formation. Here $g_s/g_n=-0.5$, $\tilde{q}=0.2$ and $V_\mathrm{in}/c_\mathrm{FDS}=0.5$. }
	\label{breatherlongtime}
\end{figure}

\section{Density waves}
\label{Density waves}
In the easy-plane phase of the ferromagnetic spin-1 BEC,  each of the two gap-less Bogoliubov  model involves both density and spin degrees of freedom~\cite{Xiaoquan22,Kawaguchi12,StamperKurn13}.   We have identified that the emitted waves by the formed breather belong to the gapless Bogoliubov mode with higher energy and the corresponding  density wave and spin wave propagate at same velocity[Fig.~\ref{densitywaves}]. 
	
\begin{figure*}[h]
		\centering
		\includegraphics[width=0.85\linewidth]{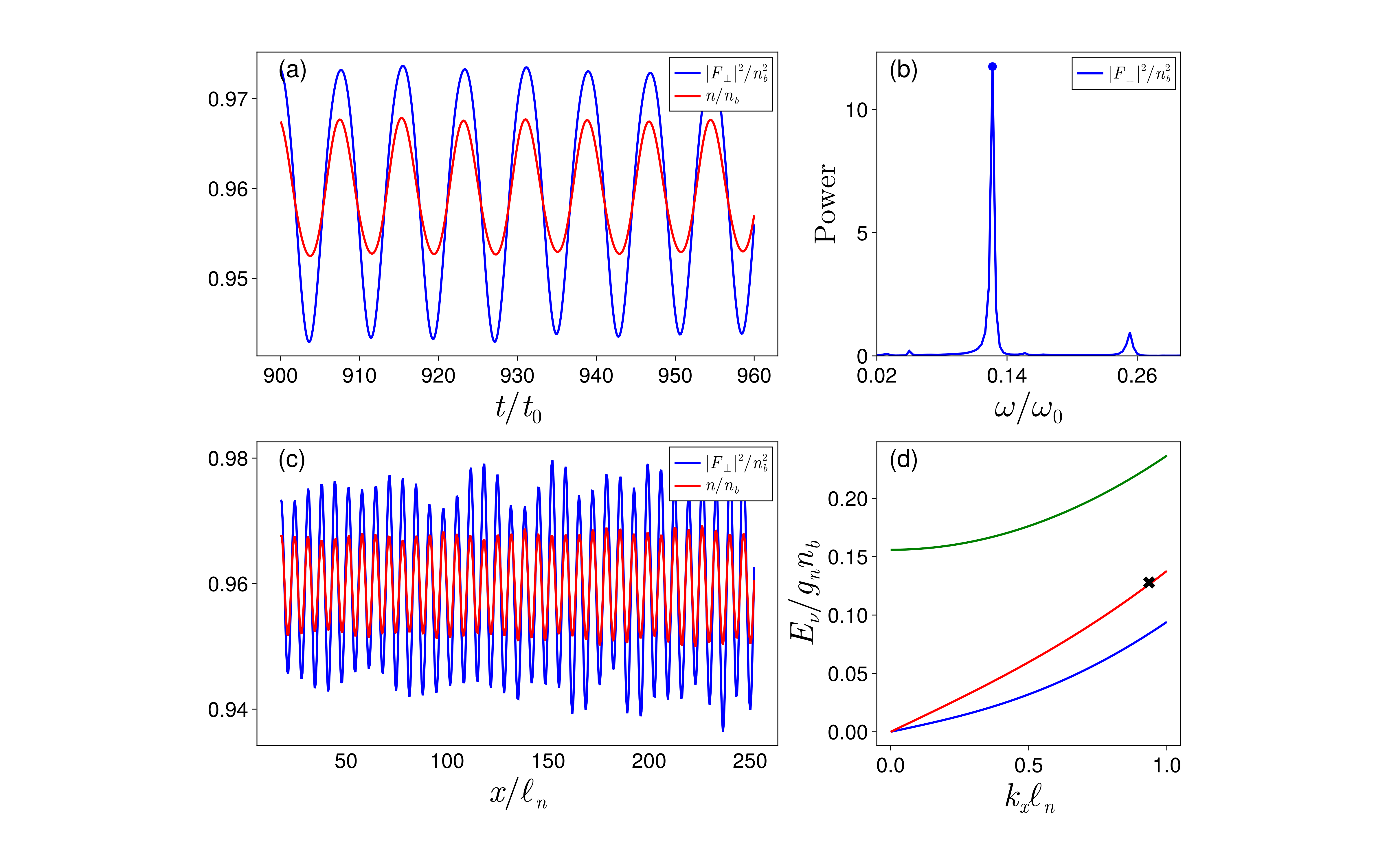}
		\caption{ (a) The time evolution of the emitted total density wave $n$ (red) and the spin wave $F_{\perp}$ (blue) at a specific spatial point. (b) The power spectrum extracted from (a).  (c) The emitted total density wave $n$ (red) and the spin wave $F_{\perp}$ (blue) at time $t/t_0 = 900$. (d) The spectrum of the Bogoliubov  modes in the easy-plane phase for $\tilde{q}=0.2$ and the location of  the emitted waves (marker)[obtained from (b) and (c)]. }
		\label{densitywaves}
\end{figure*}

\vspace{10mm}
	
It is useful to make the comparison between the Bogoliubov modes in the easy-plane phase of ferromagnetic spin-1 BECs and that in miscible two component BECs. For the two component BECs, the corresponding GP equation reads:
	\begin{eqnarray}
		\label{twocomponentGP1}
		i\hbar\partial_t\psi_1&=&-\frac{\hbar^2\nabla^2}{2M}\psi_1+g_{11}|\psi_1|^2\psi_1+g_{12}|\psi_2|^2\psi_1,\\
	i\hbar\partial_t\psi_2&=&-\frac{\hbar^2\nabla^2}{2M}\psi_2+g_{22}|\psi_2|^2\psi_2+g_{21}|\psi_1|^2\psi_2, 
		\label{twocomponentGP2}
	\end{eqnarray}
where $g_{ii}$ and $g_{ij}$ represent intra-species and inter-species interaction strength, respectively. Here we consider a miscible BEC, i.e., $g>\bar{g}$, where $g\equiv g_{11}=g_{22}$ and $\bar{g} \equiv g_{12}=g_{21}$. For given total number density $n=n_g$, the ground state wave function $\psi_g=(\psi_{1g},\psi_{2g})^{T}$ with $\psi_{1g}=\psi_{2g}=\sqrt{n_g/2}$.  Let us consider a perturbed wavefunction $\psi=\psi_g+\delta\psi$ with $\delta\psi=ue^{ik_x x-i\omega t}+v^*e^{-ik_x x+i\omega t}$.  Plugging $\psi=\psi_g+\delta\psi$ into Eqs.~\eqref{twocomponentGP1}~\eqref{twocomponentGP2}  and keeping the leading order terms,  we obtain the Bogoliubov-de Gennes (BdG) equations:
	\begin{equation}\label{BdG}	
		\hbar \omega 
		\begin{pmatrix}
			u_1 \\
			v_1 \\
			u_2 \\
			v_2
		\end{pmatrix}
		=
		\begin{pmatrix}
			\mathcal{H}_1 & g\psi_{1g}^2 & \bar{g}\psi_{1g}\psi_{2g}^* & \bar{g}\psi_{1g}\psi_{2g} \\
			-g\psi_{1g}^{*2} & -\mathcal{H}_1^* & -\bar{g}\psi_{1g}^*\psi_{2g}^* & -\bar{g}\psi_{1g}^*\psi_{2g} \\
			\bar{g}\psi_{2g}\psi_{1g}^* & \bar{g}\psi_{2g}\psi_{1g} & \mathcal{H}_2 & g\psi_{2g}^2 \\
			-\bar{g}\psi_{2g}^*\psi_{1g}^* & -\bar{g}\psi_{2g}^*\psi_{1g} & -g\psi_{2g}^{*2} & -\mathcal{H}_2^*
		\end{pmatrix}
		\begin{pmatrix}
			u_1 \\
			v_1 \\
			u_2 \\
			v_2
		\end{pmatrix},
	\end{equation}
	where $\mathcal{H}_1=\mathcal{H}_2=\hbar^2k_x^2/2M+g n_g+\bar{g} n_g/2-\mu$ and the chemical potential $\mu=\mu_1=\mu_2=(g +\bar{g})n_g/2$.  Solving Eq.~\ref{BdG}, we obtain
	\begin{equation}
		E({k_x})=\hbar \omega_{k_x}=\pm\frac{
			\sqrt{
				2 (g\pm \bar{g})  M n_g \hbar^2 k_x^2 + k_x^4 \hbar^4 }
		}{2 M}.
	\end{equation} 
	In the long-wavelength limit, i.e., $k_x\rightarrow0$, we have 
	\begin{equation}
	E (k_x)=\pm\sqrt{\frac{(g\pm \bar{g}) n_g}{2 M}}\hbar k_x=\pm c_{\pm} \hbar k_x,
	\end{equation}
where $ c_{\pm}=\sqrt{ (g\pm\bar{g} )n_g/2M }$ are the propagating velocities of the corresponding modes.  We evaluate the perturbed observables $n=|\psi_{1}|^2+|\psi_{2}|^2$ (total density) and $n_{s}=|\psi_{1}|^2-|\psi_{2}|^2$ (pseudo-spin) for the two distinct BdG modes obtained above and find  that the total density ($n$) and the pseudo-spin $n_s$ are decoupled with different propagating velocities $c_{+}$ and $c_{-}$, respectively (Fig.~\ref{2cBdG})~\cite{Timmermans98,Timmermans03,Abad13,Kamchatnov18,Shin20}.

\begin{figure}[h]
	\centering
	\includegraphics[width=0.7\textwidth]{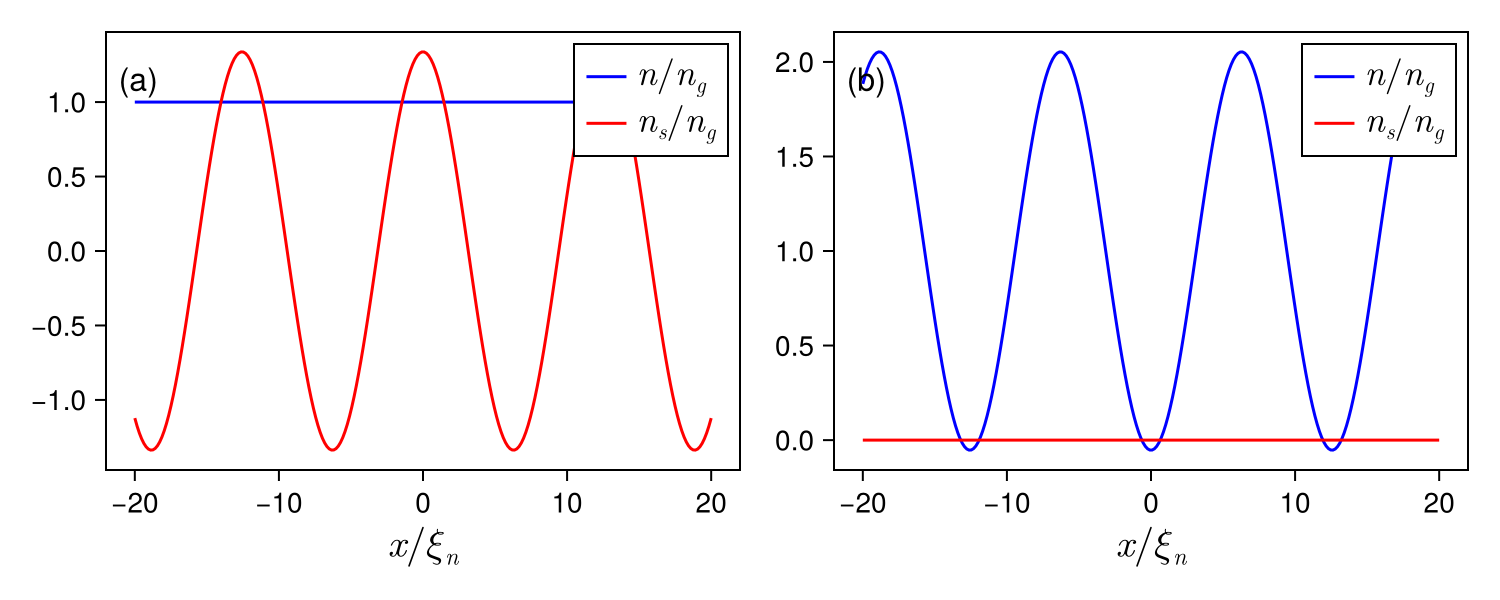}
	\caption{The perturbed observables $n$ and $n_s$ for the pseudo-magnon mode (a) and the phonon mode (b). Here $\xi_n\equiv\hbar/\sqrt{Mgn_g}$.}
	\label{2cBdG}
\end{figure}



\section{Presence and absence of destruction of the kink structure during colliding}\label{kink structure}
In the main manuscript, we claim that the collision of a type-I pair always involve destruction of the kink structure while a type-II pair only processes reflection and the destruction of the kink structure is absent.   Figure.~\ref{fig:kinkstructure} shows details of the colliding processes for type-I and type-II pairs and we can find that it is indeed the case. 

\begin{figure*}[htbp]
	\centering
	\includegraphics[width=0.49\linewidth]{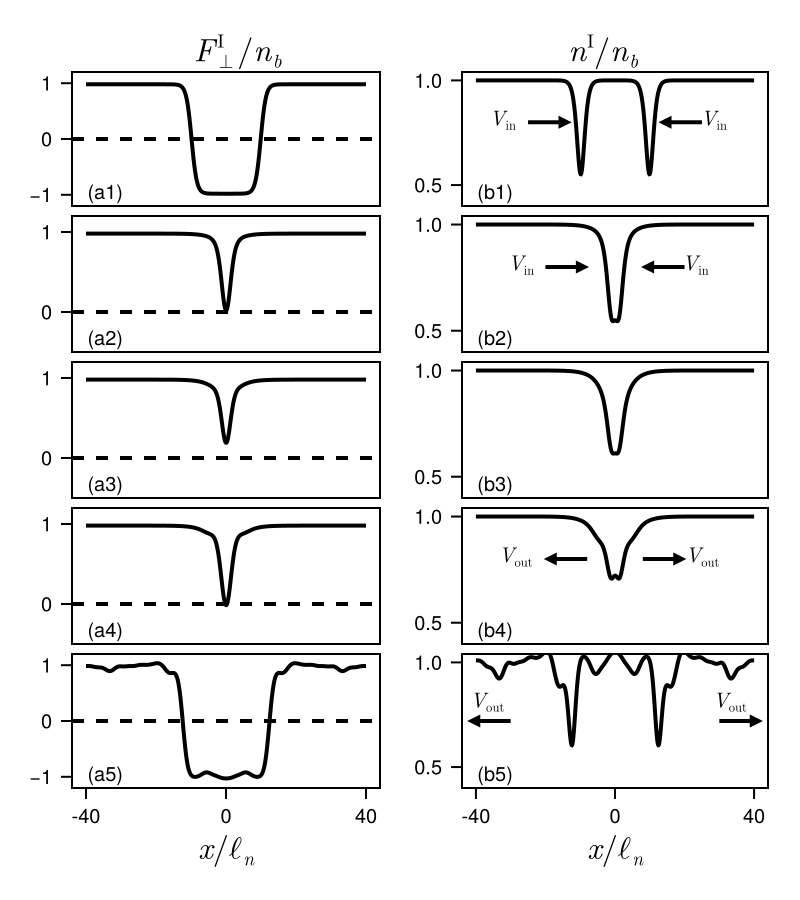}
	\includegraphics[width=0.49\linewidth]{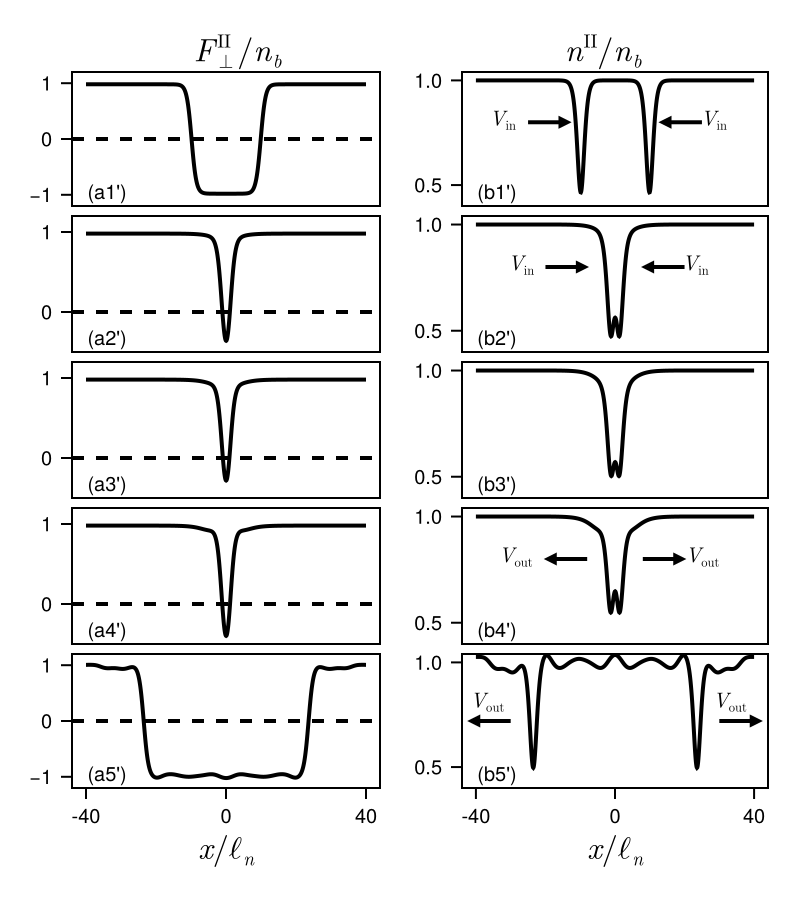}
	\caption{The profiles of the transverse magnetization $F_{\perp}$ and the total density $n$ during the type-I pair and the type-II pair collisions at the incoming velocity $V_{\rm in}/c_\mathrm{FDS}=0.9$ higher than the critical incoming velocity($V_\mathrm{c}/c_\mathrm{FDS}=0.84626$) at $t/t_0=1;69;72;74;240$. For the type-I pair collision, the topological structure in the spin order experiences destruction [a(2)-(a3)] and reproduction [(a4)-(a5)]. In contrast, for the type-II pair collision the topological structure in the spin order is well preserved [(a2')-(a4')]. Note that for the type-I pair $V_{\rm out}<V_{\rm in}$ while for the type-II pair $V_{\rm out}>V_{\rm in}$. Here $g_s/g_n=-0.5$ and $\tilde{q}=0.2$.}
	\label{fig:kinkstructure}
\end{figure*}

\section{Collisions away from the solvable parameter regime}\label{away solvable regime}
In the main manuscript, we forcus on collisions at the exactly solvable point $g_s/g_n=-1/2$.  While the characteristic features of  the collision dynamics hold in general.  Figure.~\ref{gs001} shows collisions of type-I and type-II pairs at $g_s/g_n=-0.01$ close to the value for $^{87}$Rb~\cite{Kawaguchi12, StamperKurn13}.

\begin{figure}[h]
	\centering
	\includegraphics[width=\textwidth]{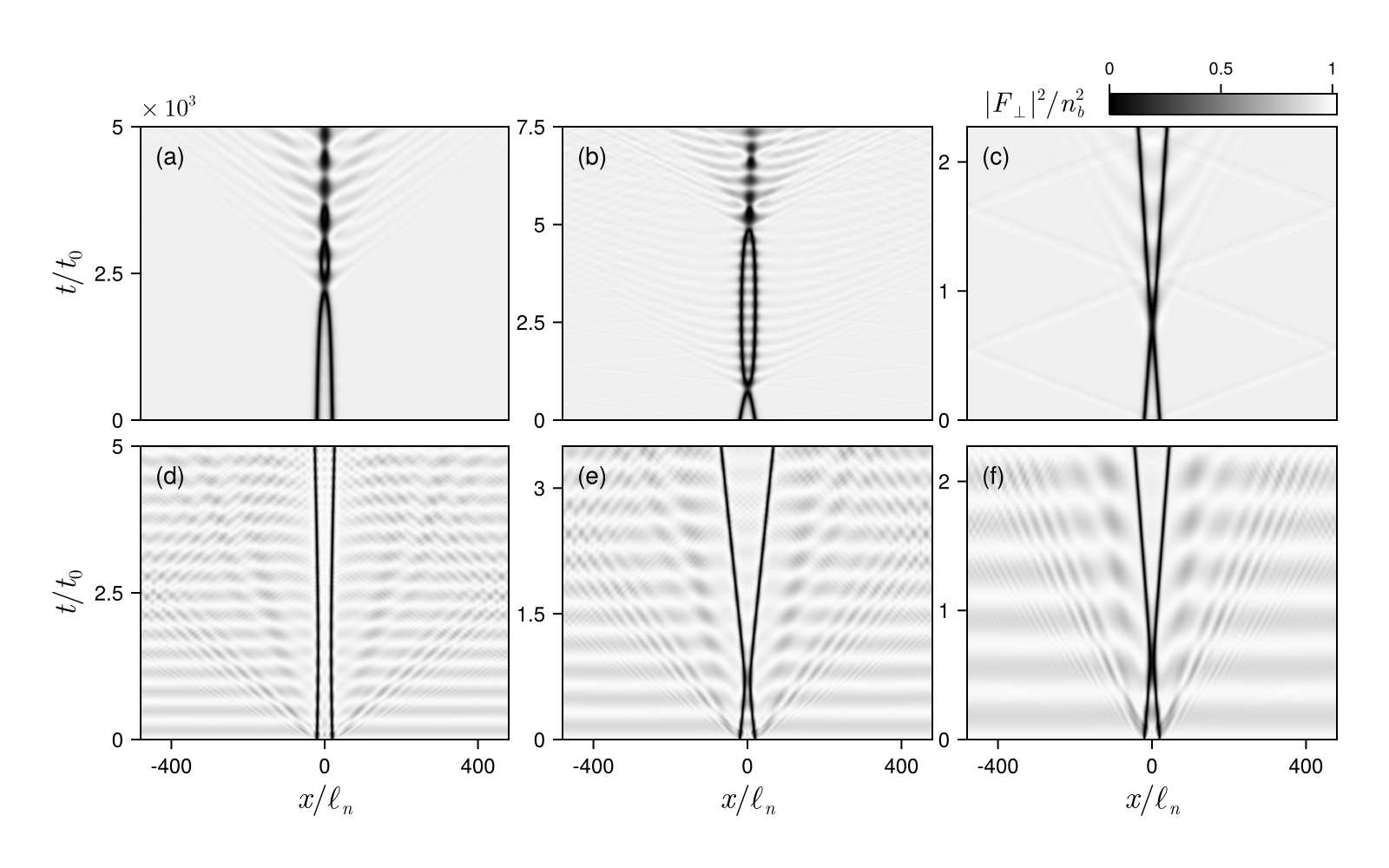}
	\caption{Collisions of type-I [(a)-(c)] and type-II [(d)-(f)] pairs at incoming velocities $V_\mathrm{in}/c=0.0003, 0.0023, 0.0028$, where $c\equiv\sqrt{g_nn_b/M}$ . Here $g_s/g_n=-0.01$ and $\tilde{q}=0.2$. }
	\label{gs001}
\end{figure}

	\twocolumngrid 
\bibliographystyle{aipnum4-1}  

	
	


	%

\end{document}